\def\year{2022}\relax
\documentclass[letterpaper]{article} 
\usepackage{aaai22}  
\usepackage{times}  
\usepackage{helvet}  
\usepackage{courier}  
\usepackage[hyphens]{url}  
\usepackage{graphicx} 
\usepackage{natbib}  
\usepackage{caption} 
\DeclareCaptionStyle{ruled}{labelfont=normalfont,labelsep=colon,strut=off} 
\usepackage{algorithm}
\usepackage{algorithmic}
\usepackage{amsthm}
\usepackage{amsmath}
\usepackage{amssymb}
\usepackage{booktabs}
\usepackage{multirow}
\usepackage{bbding}

\usepackage{newfloat}
\usepackage{listings}
\floatstyle{ruled}
\newfloat{listing}{tb}{lst}{}
\floatname{listing}{Listing}
\usepackage{color} 
\usepackage{colortbl}

\title{Image Difference Captioning with  Pre-training and Contrastive Learning}
\author {
    Linli Yao, 
    Weiying Wang, 
    Qin Jin\thanks{Qin Jin is the corresponding author.}
}
\affiliations {
    School of Information, Renmin University of China\\
    \{linliyao, wy.wang, qjin\}@ruc.edu.cn
}

\usepackage{bibentry}

\begin{document}
\maketitle
\begin{abstract}


The Image Difference Captioning (IDC) task aims to describe the visual differences between two similar images with natural language. The major challenges of this task lie in two aspects: 1) \textbf{fine-grained} visual differences that require learning stronger vision and language association and 2) \textbf{high-cost} of manual annotations that leads to limited supervised data.  
To address these challenges, we propose a new modeling framework following the pre-training-finetuning paradigm. Specifically, we design three self-supervised tasks and contrastive learning strategies to align visual differences and text descriptions at a fine-grained level. 
Moreover, we propose a data expansion strategy to utilize extra cross-task supervision information, such as data for fine-grained image classification, to alleviate the limitation of available supervised IDC data. 
Extensive experiments on two IDC benchmark datasets, CLEVR-Change and Birds-to-Words, demonstrate the effectiveness of the proposed modeling framework. The codes and models will be released at https://github.com/yaolinli/IDC.

\end{abstract}

 \section{Introduction}







Endowing machines with the ability to automatically perceive and understand visual information and express in natural language is a goal that researchers have long aspired to achieve. Image captioning~\cite{vinyals2015show, xu2015show, rennie2017self}, which aims at generating natural language description of a given image, has been one of the classic research tasks. Image Difference Captioning (IDC), which generates natural descriptions of the differences between two similar images, is a further extension of the general image captioning task, and it is more challenging~\cite{jhamtani2018learning, park2019robust, tan2019expressing}. IDC has rich potential in real world applications, such as assisting ornithologists to distinguish species with similar appearances, detecting and describing lesions automatically, and reporting salient changes in media assets and surveillance etc.



\begin{figure}[t]
    \centering
    \includegraphics[width=0.9\linewidth]{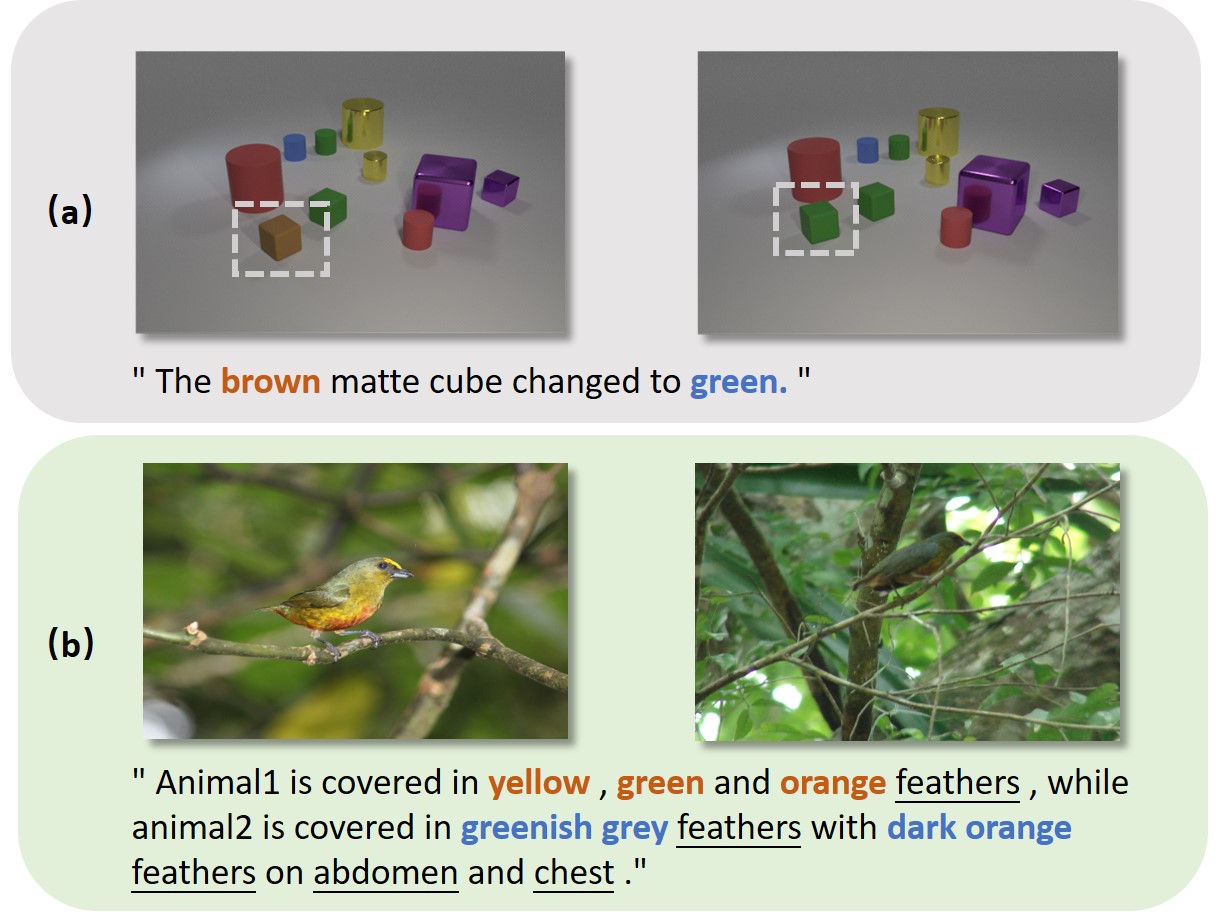}
    \caption{Image difference captioning examples. (a) an example from CLEVR-Change that involves an object change; (b) an example from Birds-to-Words that involves  detailed appearance differences of two birds.}
    \label{fig:IDC_example}
\end{figure}

Intuitively, image difference captioning involves the steps of first perception, then comparison, and finally description, which is more complex and challenging than the general image captioning task. The key challenges of image difference captioning task relate to two aspects. First, the IDC task requires \textbf{fine-grained} semantic comprehension. Different from general image captioning that describes a single image, IDC
needs to perceive the fine-grained contents of two similar images to identify the usually subtle differences. As shown in Figure~\ref{fig:IDC_example} (b), the focused differences lie in the tiny body parts of bird species (i.e.``feather'' and ``abdomen''). Moreover, the fine-grained visual differences can be very diverse in different scenarios. In case (a), we focus on the change of geometry objects in the scene, while in case (b), we only attend to the appearance of bird species regardless of the complex natural environment. 
Second, the annotation for the IDC task is particularly \textbf{high-cost}. 
Compared with general image caption annotation, it requires higher cognition load from  annotators, including first observing two images, then comparing the differences, and then using natural language to annotate these differences in order to produce the annotation in a triplet format ~\textit{(img1, img2, description)}.
\textcolor{black}{Therefore, existing manually annotated benchmark datasets are limited in data size~\cite{jhamtani2018learning,forbes2019neural,tan2019expressing}.}
There have been previous endeavors to address the above-mentioned challenges, which mainly focus on designing various attention mechanisms or improving the image features to better capture the subtle image difference based on the typical encoder-decoder structure~\cite{park2019robust, tan2019expressing, shi2020finding}.
\textcolor{black}{However, these works have not paid sufficient attention on fully interacting cross-modal fine-grained representations.} Inspired by the recent vision-language pre-training works~\cite{li2020unicoder, chen2020uniter, li2020oscar}, in this paper, we propose a new training schema for image difference captioning, which uses self-supervised learning to learn stronger associations between visual differences and language.

Our proposed training schema follows the pre-training and fine-tuning paradigm to align the visual differences with textual semantics at a fine-grained level. In the pre-training stage, we design three self-supervised tasks including Masked Language Modeling (MLM), Masked Visual Contrastive Learning (MVCL) and Fine-grained Difference Aligning (FDA). \textcolor{black}{ In MLM and MVCL tasks, we mask some parts of one modality and recover it by the other, thereby promoting the semantic interaction between visual differences and language. We replace the common feature regression objective~\cite{qi2020imagebert} on the visual side with a Noise Contrastive Estimation (NCE) loss. In the FDA task, we introduce contrastive learning strategies and construct negative contrasts to further enhance the fine-grained cross-modal association. Specifically, we carefully design three hard negatives construction strategies, namely Retrieve,  Replace, and Confuse. }
Considering the high annotation cost, we leverage extra cross-task data in our framework to learn additional background knowledge.  To be specific, we utilize datasets from  general image captioning (GIC) and fine-grained visual classification (FGVC)~\cite{ge2019weakly,liu2020filtration,dubey2018pairwise}. The GIC task provides image-text pairs and can benefit learning the alignment between images and textual descriptions. In the FGVC task, the fine-grained image labels can drive the model to learn more discriminative visual representations. Our model architecture is flexible to handle the different-form cross-task data with an image difference encoder and a multi-layer cross-modal transformer.


Extensive experiments are carried out on two benchmark datasets from different scenarios: CLEVR-Change and Birds-to-Words. Our model significantly outperforms the state-of-the-art methods on the main metrics on both benchmark datasets.
The major contributions of this work are as follows:
\begin{itemize}
\setlength{\itemsep}{-1pt}
    \item We propose a new training schema with the pre-training-finetuning paradigm for the IDC task to better align the visual difference and language by three self-supervised tasks with contrastive learning.
    \item The proposed model has a flexible structure that can utilize extra cross-task data to alleviate the limitation of supervised data due to high annotation cost.
    \item The proposed model achieves the state-of-the-art performances on the CLEVR-Change and Birds-to-Words benchmark datasets.
\end{itemize}

\section{Method}

\begin{figure*}[t]
    \centering
    \includegraphics[width=0.85\linewidth]{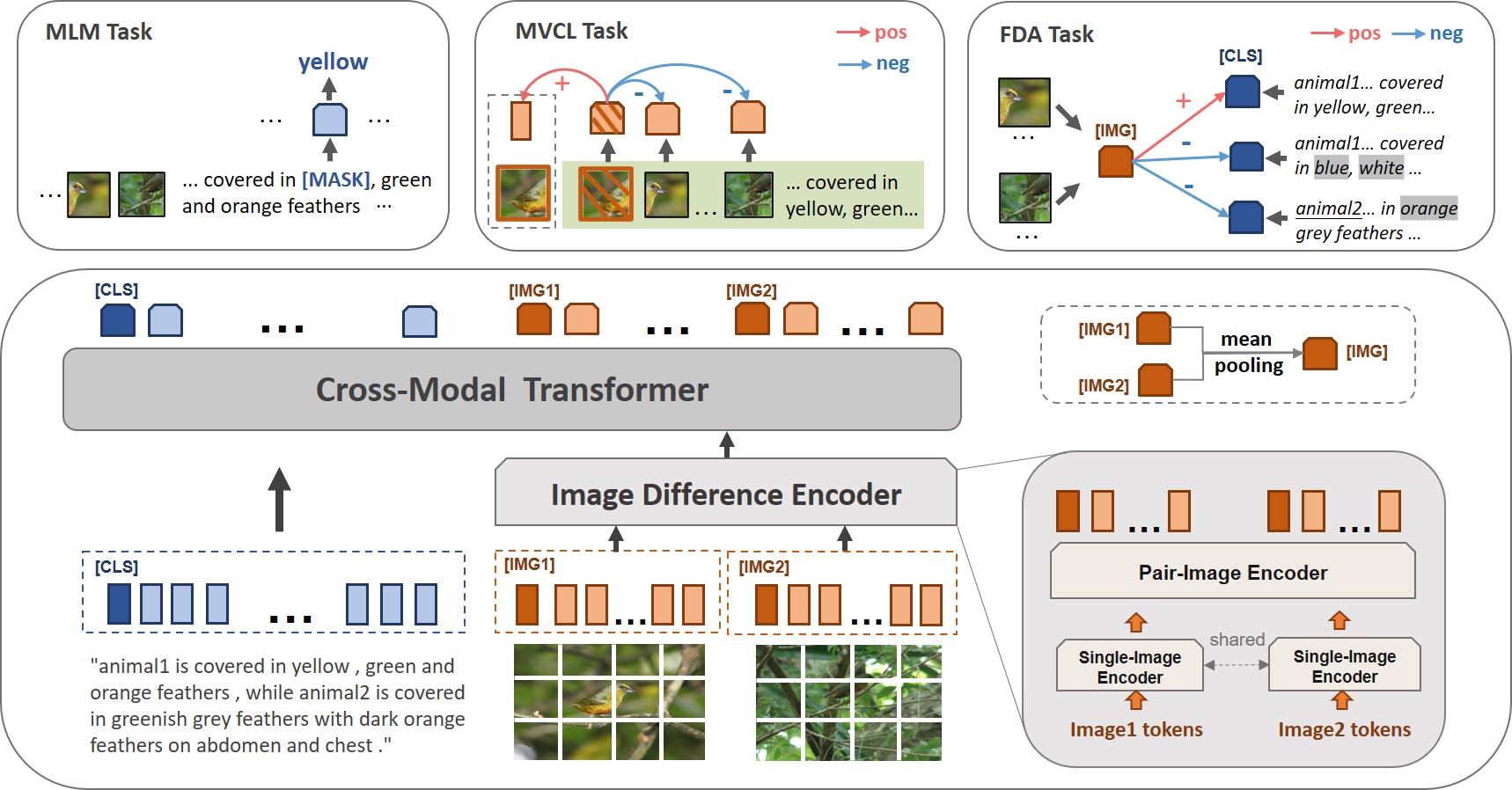}
    \caption{Overview of our proposed framework with an image difference encoder and a multi-layer cross-modal transformer (best viewed in color). Given the input in triplet format ~\textit{(img1, img2, description)}, the image pair is first fed into the image difference encoder to capture the fine-grained image differences. It consists of a single-image encoder and a pair-image encoder, which enhances the intra-image and inter-image visual representations respectively. Then the enhanced visual representations are aligned with the textual representations in the cross-modal transformer through three pre-training tasks: Masked Language Modeling (MLM), Masked Visual Contrastive Learning (MVCL) and Fine-grained Difference Aligning (FDA).  }
    \label{fig:framework}
\end{figure*}




In this section, we introduce our proposed pre-training and fine-tuning paradigm for the image difference captioning task. The overall modeling framework is illustrated in Figure~\ref{fig:framework}. It consists of an image difference encoder to capture subtle visual differences and a multi-layer cross-modal transformer to align cross-modal representations. Three pre-training tasks including  Masked Language Modeling (MLM), Masked Visual Contrastive Learning (MVCL) and Fine-grained Difference Aligning (FDA) are designed to take most advantage of the given data. To handle the problem of limited supervised IDC data, we 
expand cross-task data in our flexible framework.

\subsection{Model Architecture}

\textbf{Input Representation}
We define the input of the IDC task, which contains a pair of images and a text description, as ${\{V^{(1)}, V^{(2)}, T\}}$.
For the text description, we tokenize each word in the sentence and convert it to word embedding trained from scratch, denoted as:
\begin{small}
\begin{equation}
    T = {\{\text{[CLS]},\text{[BOS]},w_0,\ldots,w_M,\text{[EOS]}\}}
\end{equation}
\end{small}
where the special token [CLS] is added to capture the global semantics of the sentence. Following previous works~\cite{park2019robust, tan2019expressing, forbes2019neural}, we use pre-trained ResNet101 \cite{resnet101} to extract grid features for the two images, denoted as:
\begin{small}
\begin{align}
    V^{(1)} = \{ \text{[IMG1]}, v^{(1)}_0,\ldots,v^{(1)}_i,\ldots,v^{(1)}_N\} 
    \\
    V^{(2)} = \{ \text{[IMG2]}, v^{(2)}_0,\ldots,v^{(2)}_i,\ldots,v^{(2)}_N\}
\end{align}
\end{small}
Similar to text representations, we also add two special tokens [IMG1] and [IMG2] to extract global image semantics respectively. A linear layer is applied to keep the visual feature dimension the same as the word embedding.

To explicitly indicate the positions of tokens inside each modality, we add fixed positional embeddings~\cite{vaswani2017attention} to each token as well. In addition, we employ type embeddings to distinguish whether a token belongs to  ${V^{(1)}}$, $V^{(2)}$, or $T$.

\noindent \textbf{Image Difference Encoder}
Based on the intuitive cognition process of image difference captioning, observing two images and then comparing them, we design an image difference encoder, which consists of a single-image encoder and a pair-image encoder.
The single-image encoder $\mathcal{F}_{\text{sing}}$ takes visual tokens from an image as input to embed the semantics of different regions in the image. Then embeddings of the two images are fed into the pair-image encoder $\mathcal{F}_{\text{pair}}$, which can interact the visual semantics between two images and implicitly learn to locate image differences.
We use the transformer architecture for both $\mathcal{F}_{\text{sing}}$ and $\mathcal{F}_{\text{pair}}$. 
\begin{small}
\begin{equation}
    \widetilde{V}^{(1)},\widetilde{V}^{(2)} = \mathcal{F}_{\text{pair}}\left(\mathcal{F}_{\text{sing}}(V^{(1)}),\mathcal{F}_{\text{sing}}(V^{(2)})\right)
\end{equation}
\end{small}

\noindent \textbf{Cross-modal Transformer}
We utilize self-attention based multi-layer transformer for the cross-modal encoder to align context between visual and textual modalities. 
\begin{small}
\begin{equation}
    \hat{V}^{(1)},\hat{V}^{(2)}, \hat{T} = \mathcal{F}_{\text{cross}}
    \left(\widetilde{V}^{(1)},\widetilde{V}^{(2)},T\right)
\end{equation}
\end{small}

\subsection{Pre-training Tasks}
We design three pre-training tasks to enhance the fine-grained alignment between image differences and captions so that to learn better feature representations. 

\textbf{Masked Language Modeling (MLM)}
For textual side, we apply Masked Language Modeling to promote the context mapping from vision to language, following existing VLP works \cite{chen2020uniter,huang2021seeing}. The objective of the MLM task is to predict the masked word $w_{m}$ based on the surrounding unmasked words $w_{\backslash m}$ and visual difference context $\{\widetilde{V}^{(1)}, \widetilde{V}^{(2)}\}$. Similar to BERT, we mask 15\% of the input word tokens in which 80\% are replaced with a special token [MASK], 10\% with random words and 10\% unchanged. The masked hidden output of cross-modal transformer are fed into a classifier to predict the original words. We formulate the training objective of MLM task as:
\begin{small}
\begin{equation}
    \mathcal{L}_{\text{MLM}}= \mathbb{E}_{V,T \in D} \left[ - \log P_{\theta}\left(w_{m} \mid w_{\backslash m}, \widetilde{V}^{(1)}, \widetilde{V}^{(2)} \right) \right]
\end{equation}
\end{small}
where $D$ denotes the entire training set and $\theta$ is the model parameters to be learned.

\vspace{8pt}
\textbf{Masked Visual Contrastive Learning (MVCL)}
Similar to the MLM task, we also apply masking and recovering strategy on the visual side. 
Since visual representation is continuous and high-dimensional, the goal of the MVCL task is to reconstruct the masked image features according to the difference caption and the remaining visual semantics. Specifically, we mask 15\% input image features and replace the masked features with zero vectors. 
Note that we only mask features from one image each time to ensure the masked content can be recovered by the other two in triplet ${\{V^{(1)}, V^{(2)}, T\}}$. The general objective of MVCL task is:
\begin{small}
\begin{equation}
    \mathcal{L}_{\text{MVCL}}=\mathbb{E}_{V,T \in D} f_{\theta}\left(v_{m} \mid v_{\backslash m},T \right)
\end{equation}
\end{small}
where $V = \{V^{(1)}, V^{(2)}\}$. Inspired by the video language pre-training work \cite{Luo2020UniVL,li2020hero}, we introduce contrastive learning and use a NCE loss~~\cite{sun2019learning} to define $f_{\theta}\left(v_{m} \mid v_{\backslash m},T \right)$ as:
\begin{small}
\begin{equation}
    -\log 
        \frac{ \exp \left(d(v_{m}, v_{m}^{+}) / \tau_{1} \right)}
        {\exp \left(d(v_{m}, v_{m}^{+}) / \tau_{1} \right) + 
        \sum_{v^{\prime} \in \mathcal{N}(v_m)} \exp \left(
            d(v_{m}, v^{\prime}) / \tau_{1}
        \right)}
\end{equation}
\end{small}
where $d(.)$ denotes the cosine similarity, $\tau_{1}$ is the temperature hyper-parameter and $v_{m}^{+}$ denotes the original  image feature of $v_{m}$ before masking. We define the unmasked image features in the batch as negative samples $\mathcal{N}(v_m)$. 
The contrastive loss pushes the model to identify the positive sample $v_{m}^{+}$ of $v_{m}$ from negative samples $\mathcal{N}(v_m)$ in the batch, which enforces the reconstructed image representations of $v_{m}$ to be more discriminative.



\begin{figure}[t]
    \centering
    \includegraphics[width=\linewidth]{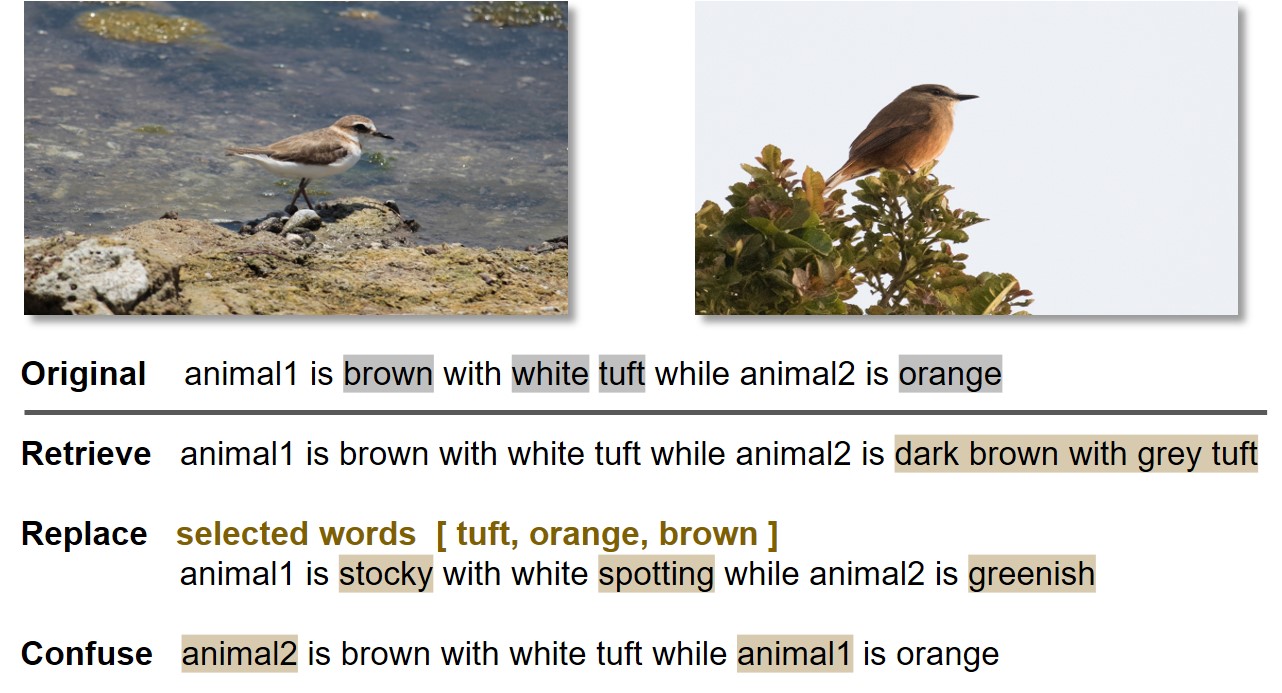}
    \caption{An example of constructed negative sentences in FDA task on Birds-to-Words Dataset.}
    \label{fig:neg_samples}
\end{figure}

\textbf{Fine-grained Difference Aligning (FDA)}
\textcolor{black}{To explicitly bridge visual and textual modalities in a more fine-grained way, we introduce contrastive learning and construct hard negative samples.
To be specific, we rewrite the original difference caption in three ways, as illustrated in Figure~\ref{fig:neg_samples}:}


\begin{itemize}
    \item \textbf{Retrieve}  For each triplet sample $\{V^{(1)},V^{(2)}, T\}$, we use TF-IDF similarity to retrieve the most similar difference descriptions $\{T^{-}\}$ from other samples for $T$ and consider them as hard negative samples.
    
    \item \textbf{Replace}  We replace the most important difference-related words in the caption to facilitate fine-grained alignment. Empirically, we observe that the attribute words in the caption are more related to the difference (e.g. ``grey'', ``beak''). Therefore, we first annotate adjectives and nouns in each sentence using Stanford CoreNLP tool. Then we sort the annotated words by TF-IDF score to measure their importance. Top K (50\%) annotated words are selected and replaced by other words with the same POS-Tags that are randomly sampled from pre-defined vocabularies. 
    \item \textbf{Confuse} If we change the subject in the difference description, the semantics of the sentence will totally change while the structure remains similar. For example, changing ``\textit{\underline{animal1} has a longer tail than \underline{animal2}}'' to ``\textit{\underline{animal2} has a longer tail than \underline{animal1}}''. We achieve this by changing the subjects between sentences or switching the subject and object within a sentence.

\end{itemize}

Based on the positive sample $(V, T^{+})$ and constructed 
negative samples  $(V, T^{-})$, we employ a contrastive loss to define the training objective $ \mathcal{L}_{\text{FDA}} = \mathbb{E}_{V,T \in D}\left[ -\log \text{NCE}(V,T) \right]$, in which $\text{NCE}(V,T)$ is:
\begin{small}
\begin{equation}
            \frac{\exp \left(d\left(V, T^{+}\right) / \tau_{2}\right)}
            { \exp \left(d\left(V, T^{+}\right) / \tau_{2}\right) + \sum_{T^{-} \in \mathcal{N}_T} \exp \left(d\left(V, T^{-}\right) / \tau_{2}\right)}
\end{equation}
\end{small}
where $d(.)$ denotes the cosine similarity, $\tau_{2}$ is a hyperparameter and $\mathcal{N}_T$ is the negative text set. We take the average of special token [IMG1] and [IMG2] from the output as the global visual representation,  and the special token [CLS] from the output as the textual representation.

In the pre-training stage, we use bi-directional attention mask, thus the  special tokens [IMG1], [IMG2], [CLS] can learn global joint representations from images and text. We train different pre-training tasks by alternating batch with different ratio.

\vspace{-1pt}
\subsection{Finetuning and Inference}
After pre-training, we fine-tune our model to generate difference captions based on effective visual semantics across two similar images. Similar to \cite{li2020oscar}, we adapt the MLM task in the fine-tuning stage. 
Different from the pre-training stage, we employ a uni-directional attention mask to restrict the attention on the text side, which means that each word can only attend to its previous words. While the attention on the visual side does not change and each word can attend to all visual tokens. In this way, we can keep fine-tuning and pre-training as consistent as possible while adapting the model to sentence generation as much as possible.


During the inference stage, the model generates the difference caption word by word based on visual difference semantics. 
The embeddings of all visual features $\{V^{(1)}, V^{(1)}\}$ and the special token [CLS] are used as input. The start token [BOS] with a [MASK] token are then fed to the model to trigger the sentence generation. The model samples a word $w_0$ from the vocabulary based on the likelihood output of the [MASK] token. At step $t$, the [BOS] token, previous generated words $w_{\textless t}$ and a new [MASK] token are fed to the model to generate next word $w_{t}$. The model can thus generate the whole sentence until [EOS] is sampled. 


\vspace{-3pt}
\subsection{Expansion of Cross-task Data}
To alleviate the limitation of available IDC triplet data, we propose a data expansion strategy to utilize extra cross-task data. Specifically, we expand data from general image captioning and fine-grained visual classification tasks. 

\textcolor{black}{\textbf{General image captioning(GIC)} task aims to describe a single image with sentences. The GIC data is in the \textit{(image, text)} format, which can facilitate the model to learn preliminary cross-modal alignment. In order to adapt to the triplet data format of the IDC task, we pad an empty image with zero vectors to form the pseudo triplet. 
In cross-modal transformer, the padded vectors will not be involved in self-attention calculation by a special attention mask. The pseudo triplet input can naturally adapt to three pre-training tasks. Note that in the MVCL task, we only mask tokens from the real image.} 

\textcolor{black}{
\textbf{Fine-grained visual classification(FGVC)} is a challenging task to classify images with subtle inter-class differences. The slightly different images with different class labels can enhance our image difference encoder to learn more discriminative visual representations. 
Specifically, we construct image pairs \textit{(img1, img2)} by random sampling, half of which are from the same class label, and the other half are not.
Each single image (i.e. \textit{img1} or \textit{img2}) in  the obtained pairs is used to refine the single-image encoder $\mathcal{F}_{\text{sing}}$ with a fine-grained classification objective and a contrastive loss~\cite{he2021transfg}. 
Then the dual visual representations are fed to the pair-image encoder $\mathcal{F}_{\text{pair}}$ to optimize it with a matching loss which verifies whether the two images \textit{(img1, img2)} are from the same class. The matching loss enhances the ability of the difference encoder to compare two similar images. 
}
Note that we select cross-task datasets with similar semantic domain as the IDC benchmark dataset, so they can  provide more similar background knowledge for multi-modal representation learning and cross-modal alignment.

\section{Experiments}


We evaluate our model on two IDC benchmark datasets from different domains including CLEVR-Change~\cite{park2019robust} and Birds-to-words~\cite{tan2019expressing}. 

\begin{table}[t]
\begin{center}
\small
\begin{tabular}{l|ccc >{\columncolor[gray]{0.9}} c}
\toprule
Model 
&B4        & M       & C(D)    & \textbf{R} \\ \midrule
Neural Naturalist~\shortcite{forbes2019neural}  & 22.0 & - & 25.0  & 43.0          \\
Relational Speaker~\shortcite{tan2019expressing} & 21.5 & 22.4 & 5.8 & 43.4  \\
DUDA~\shortcite{park2019robust}  & 23.9 & 21.9 & 4.6 & 44.3   \\ 
L2C~\shortcite{yan2021l2c} & 31.3 & - & 15.1 & 45.3 \\
L2C(+CUB)~\shortcite{yan2021l2c} & \textbf{31.8} & - & 16.3 & 45.6   \\ \midrule
Ours &28.0 &23.1 &18.6  &48.4   \\
Ours(+Extra Data) &31.0 &\textbf{23.4} &\textbf{25.3}  &\textbf{49.1}  \\
\bottomrule
\end{tabular}
\end{center}
\caption{Comparison with state-of-the-art models on \textbf{Birds-to-Words} dataset. B4, M, R, and C(D) are short for BLEU-4, METEOR, ROUGE-L and CIDEr(D). The main metric ROUGE-L on this dataset is highlighted. 
}
\label{tab:SOTA_Bird}
\end{table}

\subsection{Experimental Settings}
\subsubsection{Benchmark Datasets}
CLEVR-Change dataset~\cite{park2019robust} is automatically built via the CLEVR engine and describes scene changes of geometry objects with clean background.  It has 67,660, 3,976 and 7,970 image pairs for training, validation and test split respectively. Each image pair is annotated with 6.2 captions on average.
Birds-to-Words dataset~\cite{tan2019expressing} describes the fine-grained difference of various bird species collected in the wild. It has 4,860 image pairs and each pair corresponds to  3.31 annotated captions on average. 

\subsubsection{Cross-task Datasets}
CUB~\cite{cub} serves as a single image captioning dataset. It contains 11,788 images of 200 bird species and each image is annotated with 10 captions. We use the split of  8855 training images and 2933 validation images following~\cite{yan2021l2c}. 
NABirds~\cite{nabirds}  is a fine-grained visual classification dataset which contains 48,562 images of North American birds with 555 categories.

\subsubsection{Metrics}
We report results on the standard image captioning metrics including BLEU~\cite{bleu}, METEOR~\cite{meteor}, ROUGE-L~\cite{rouge} and CIDEr(CIDEr-D)~\cite{cider} following previous works. 
In addition, we particularly emphasize the main metric that has been commonly recognized in previous works, which refers to: CIDEr for CLEVR-Change and ROUGE-L for Birds-to-Words.

\subsubsection{Implementation Details}
We extract grid image features in the shape of (7,7,2048) using the pre-trained ResNet101~\cite{resnet101} and flatten it to a feature sequence in the shape of (49, 2048). The word embedding  is learned from scratch and its dimension is 512. For the cross-modal transformer, the hidden size is 512, the attention head is 8, and the layer number is 2 for Birds-to-Words and 3 for CLEVR-Change. We set $\tau_1$, $\tau_2$ in contrastive learning to 1. In FDA task, we rewrite 6 negative sentences for each image pair, among which retrieve:replace:confuse=2:2:2. For CLEVR-Change, we sample the batch from the three pre-training tasks with ratio of MLM:MVCL:FDA=8:1:2. We pre-train the model with 8K warm-up steps and 250K iterations in total.  For Birds-to-words, the ratio of pre-training tasks is MLM:MVCL:FDA=9:1:2. The warm-up steps are 4K and total training steps are 50K.
In the pre-training stage, we apply Adam~\cite{kingma2014adam} optimizer with learning rate 1e-4. In the finetuning stage, the learning rate is set as 3e-5. Early-stop is applied on the main metric to avoid overfitting. The sentence is generated with greedy search in inference. More details can be found in the supplementary material.  


For CLEVR-Change, we notice that half of the data samples belong to a \textit{distractor} type, which means that
there are only non-semantic differences between the images, e.g. angle, zoom, or illumination changes. Accurately distinguishing semantic changes from distractors has great impact on model performance. We therefore jointly train  a distractor judging task in the finetuing stage. Specifically, we concatenate the representation of special token [IMG1] and [IMG2] and feed it to a binary classifier to judge whether the visual change is a distractor or not. 

\begin{table}[t] 
\begin{center}
\small
\begin{tabular}{l|ccc >{\columncolor[gray]{0.9}} c}
\toprule
\textbf{Model}  
 & B4        & M        & R       & \textbf{C}     \\ \midrule
Capt-Dual-Att~\shortcite{park2019robust} &43.5 &32.7 &- &108.5 \\
DUDA~\shortcite{park2019robust} & 47.3 & 33.9 & - & 112.0 \\
VAM~\shortcite{shi2020finding} & 50.3 & 37.0 & 69.7 & 114.9 \\
VAM+~\shortcite{shi2020finding} & \textbf{51.3} & \textbf{37.8} & 70.4 & 115.8 \\
IFDC~\shortcite{tmm} & 49.2 & 32.5 & 69.1 & 118.7 \\
DUDA+Aux~\shortcite{2021Auxiliary} & 51.2 &37.7 &70.5 & 115.4 
\\\midrule
Ours &51.2 &36.2 &\textbf{71.7} &\textbf{128.9}   \\
\bottomrule
\end{tabular}
\end{center}
\caption{Comparison with state-of-the-art models on \textbf{CLEVR-Change} dataset. B4, M, R, and C are short for BLEU-4, METEOR, ROUGE-L and CIDEr. The main metric CIDEr on this dataset is highlighted.
}
\label{tab:SOTA_CLEVR}
\end{table}

\subsection{Comparison with the state-of-the-arts}
\subsubsection{Results on Birds-to-Words}
We evaluate our method on Birds-to-Words compared with other state-of-the-art models,  including Neural Naturalist~\cite{forbes2019neural}, Relational Speaker~\cite{tan2019expressing}, DUDA~\cite{park2019robust} and L2C/L2C(+CUB)~\cite{yan2021l2c}. As shown in Table~\ref{tab:SOTA_Bird}, our model without extra cross-task data has achieved significant improvement on the main metric ROUGE-L (from 45.6 to 48.4), even outperforming L2C with extra CUB data. The extra cross-task data further  improves the model performance on all metrics, which demonstrates that more data from similar domain can provide beneficial background knowledge. Our model with extra data achieves the new state-of-the-art performance on METEOR, CIDEr-D and ROUGE-L.

\subsubsection{Results on CLEVR-Change}
We compare the proposed model with other state-of-the-art models with typical encoder-decoder structure, including Capt-Dual-Att~\cite{park2019robust}, DUDA~\cite{park2019robust},  VAM/VAM+~\cite{shi2020finding}, IFDC~\cite{tmm} and DUDA+Aux~\cite{2021Auxiliary}. Table~\ref{tab:SOTA_CLEVR} shows that our proposed model significantly outperforms all previous models on the main metric CIDEr, boosting previous best CIDEr score of 118.7 to 128.9. Compared with models employing well-designed attention mechanism or visual encoder, our architecture is more straightforward and flexible which achieves improved performance via effective self-supervised tasks. We also evaluate results on breakdown change types as shown in Table~\ref{tab:CLEVR_type}. Our model  achieves better CIDEr scores on three types including Texture, Add and Distractor due to the fine-grained alignment across modalities in pre-training. 

Since CLEVR-Change and Birds-to-Words are from entirely different domains, the experiment results indicate that our model can robustly capture diverse  visual differences.

\begin{table}[t]
\begin{center}
\small
\begin{tabular}{l|cccccc}
\toprule
\textbf{Model} 
 & C  & T  & M  & A & D & DI    \\ \midrule
DUDA~ & 120.4 & 86.7 & 56.4 & 108.2 &103.4 &110.8 \\
VAM+~ & 122.1 & 98.7 & 82.0 & 126.3 &115.8 &122.6 \\
IFDC~ & \textbf{133.2} & 99.1 & \textbf{82.1} & 128.2 &\textbf{118.5} &114.2 \\
\midrule
Ours  & 131.2 & \textbf{101.1} & 81.7 & \textbf{133.3} &116.5 &\textbf{145.0}  \\
\bottomrule
\end{tabular}
\end{center}
\caption{Breakdown CIDEr performance on different type of changes of \textbf{CLEVR-Change} Dataset: C(Color), T(Texture), M(Move), A(Add), D(Drop) and DI(Distractor).  
}
\label{tab:CLEVR_type}
\end{table}

\begin{table}[t]
\begin{center}
\small
\begin{tabular}{l|c|ccc >{\columncolor[gray]{0.9}} c}
\toprule

\textbf{Pre-training Tasks}   & \textbf{DE}   & B4        & M        & R  & \textbf{C}    \\ 
 \midrule
1    None & \checkmark &32.7 &27.7 &57.2 &89.8   \\
2     \text{MLM } & \checkmark &36.7  &28.2   &60.9   &94.9   \\
3     \text{MLM +  \text{MVCL} } & \checkmark &50.3 &37.6 &70.6 &119.7\\
4     \text{MLM  +  \text{MVCL} + \text{FDA}} & \checkmark  &51.2 &36.2 &71.7 &128.9   \\
5    \text{MLM +  \text{MVCL} + \text{FDA}} & \XSolidBrush  &49.2 &35.8 &68.8 & 107.9   \\
\midrule
6  w/o Distractor Judging  & \checkmark  &49.8 &36.9 &69.2 & 123.5   \\

\bottomrule
\end{tabular}
\end{center}
\caption{Ablation study results on CLEVR-Change dataset. \textbf{DE} is short for Image \textbf{D}ifference \textbf{E}ncoder module in our model. B4, M, R, and C are short for BLEU-4, METEOR, ROUGE-L and CIDEr. The main metric CIDEr on this dataset is highlighted. 
}
\label{tab:Ablation}
\end{table}

\subsection{Results and Analysis}

\subsubsection{Pre-training for IDC}
We first validate the effectiveness of three self-supervised tasks on CLEVR-Change dataset in Table~\ref{tab:Ablation}. Line 1 shows the result from the baseline without pre-training, which  can be considered as performing the finetuning stage only. The CIDEr score is obviously low in this case  without any further interaction across modalities. When more self-supervised tasks are added in the pre-training stage (Line 2$\sim$4), the model performance increases accordingly, which proves the effectiveness of our pretraining-finetuning paradigm. It is worth noting that  adding MVCL task brings large performance improvement (CIDEr from 94.9 to 119.7), which benefits from the mask-recover schema  on the visual side. 
The FDA task further improves the CIDEr score from 119.7 to 128.9, which indicates that contrastive learning with well-designed hard negative samples helps fine-grained cross-modal alignment. 
Furthermore, we analyze the impact of the image difference encoder module in Line 5, where the results show that the performance drops significantly without the image difference encoder. It indicates that the image difference encoder can effectively learn to capture subtle visual differences. \textcolor{black}{Compared to Line 4, Line 6 is the result without distractor judging task in the finetuning stage, which proves that this task can help model distinguish semantic changes from distractors and improve the model performance.}

\subsubsection{Cross-task Data Usage}
We evaluate the effectiveness of cross-task data usage on Birds-to-Words in Table~\ref{tab:in_domain}. 
CUB is a general image captioning (GIC) dataset that describes bird species, and NABirds is a fine-grained bird image classification dataset. The experiment results show that leveraging CUB can bring more performance improvement because it promotes learning the alignment between images and text descriptions, while NABirds only enhances the visual side. A combination of both cross-task datasets  provides more background knowledge and thus achieves the best result. 

\begin{table}[t]
\begin{center}
\small
\begin{tabular}{l|ccc|ccc >{\columncolor[gray]{0.9}}c}
\toprule
\textbf{Model} 
&B2W  &CUB  &NAB 
 & B4  & M   & C(D) & \textbf{R}  \\ \midrule
\multirow{2}{1cm}{L2C} & \checkmark  & & & 31.3 & - & 15.1   & 45.3 \\
 & \checkmark  &\checkmark   &           & 31.8 & - & 16.3  & 45.6 \\

\midrule
\multirow{4}{1cm}{Ours}
& \checkmark  &            &           &28.0 &23.1 &18.6    &48.4 \\
& \checkmark  &\checkmark  &           &29.3 &23.1 &23.8   &48.5 \\
& \checkmark  &            &\checkmark &27.5 &23.3 &21.9   &48.5 \\
& \checkmark  &\checkmark  &\checkmark &31.0 &23.4 &25.3   &49.1 \\
\bottomrule
\end{tabular}
\end{center}
\caption{Model performance on Birds-to-Words(B2W) dataset using two  cross-task dataset including CUB and NABirds(NAB). B4, M, R, and C(D) are short for BLEU-4, METEOR, ROUGE-L and CIDEr-D. The main metric ROUGE-L on this dataset is highlighted. 
}
\label{tab:in_domain}
\end{table}

\begin{figure*}[t]
    \centering
    \includegraphics[width=0.96\linewidth]{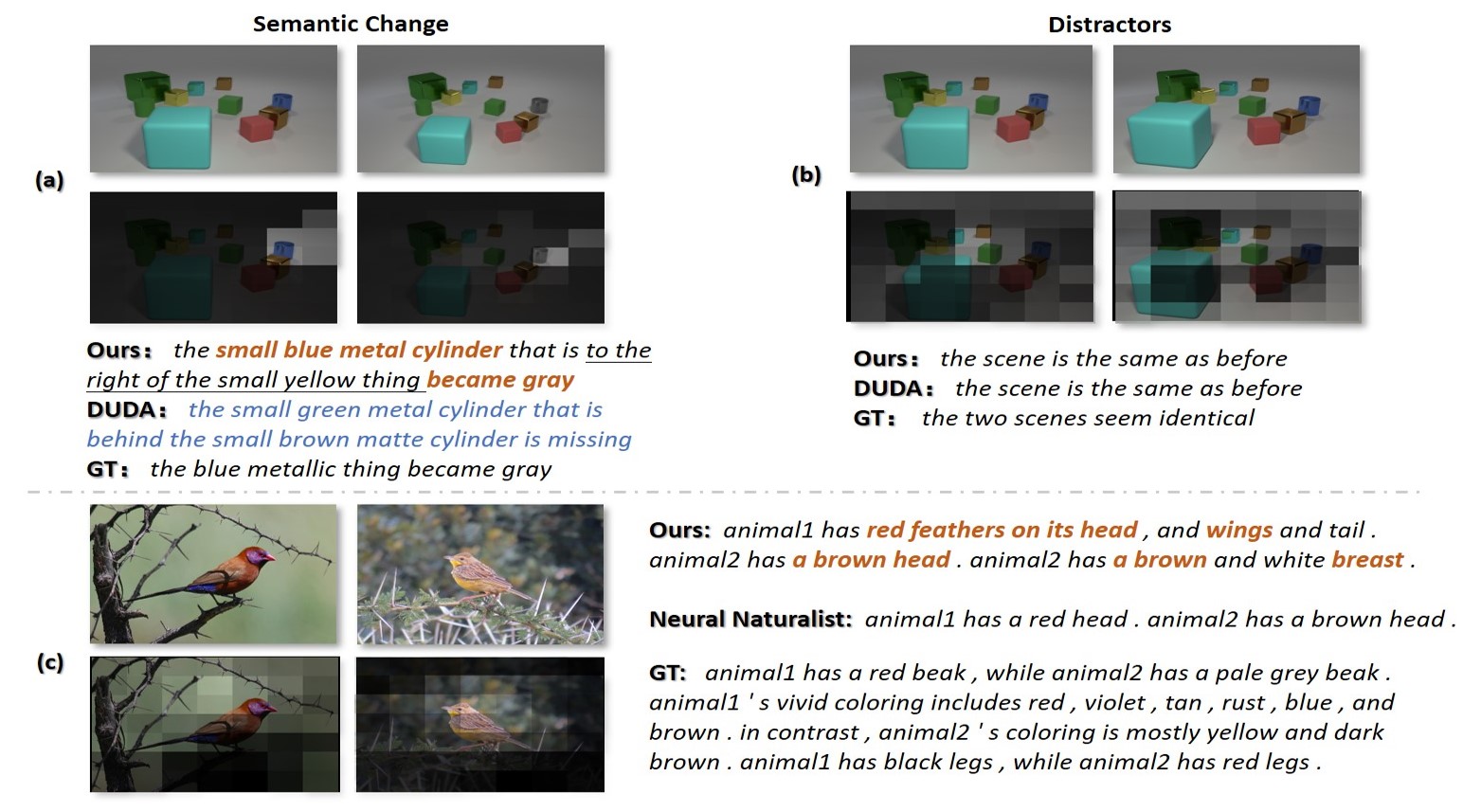}
    \caption{ Visualization of generated cases (best viewed in color). The first block illustrates two cases from CLEVR-Change. Case (a) involves semantic changes while case (b) is a distractor with only viewpoint shift. For each case, the first row is the original image pair and the second row is the corresponding attention maps. When the attention score is higher, the region is brighter. Case (c) is from Birds-to-Words and it involves fine-grained bird appearance difference with complex background. Orange bold words are correct generated  difference-related words while blue words are wrong. }
    \label{fig:case}
\end{figure*}

\subsubsection{Qualitative Results}
\textcolor{black}{
We visualize some cases in Figure~\ref{fig:case}. The top two cases are from CLEVR-Change. (a) The attention maps show that our proposed model can correctly capture the fine-grained semantic changes. (b) Our model can accurately identify the distractor scenario.
The bottom cases (c) are from Birds-to-Words, which demonstrate that our model can capture the fine-grained bird appearance difference regardless of the complex background. We also observe that the proposed model tends to repeat the same differences or ignore the conflicts in the generated sentences, which could be further addressed in our future work.
More visualizations and analysis are available in the supplementary material. }

\section{Related Works}
\label{sec:Related_Work}

\noindent \textbf{Image Difference Captioning}
More challenging than general image captioning~\cite{vinyals2015show, xu2015show, jiang2018learning, gu2018stack, zhao2020memcap, fei2021partially}, the image difference captioning task promotes the research of vision and language to achieve a more fine-grained understanding. 
Recently, several benchmark datasets of IDC have emerged \cite{jhamtani2018learning, park2019robust, tan2019expressing, forbes2019neural}, which are collected from different domains with different focus.  
However, most of the datasets are limited in scale due to the high annotation cost.
Prior works mainly focus on employing well-designed attention mechanisms~\cite{park2019robust,tan2019expressing,shi2020finding} or improving  visual features~\cite{yan2021l2c,tmm} to better capture the subtle visual difference. Besides, \cite{2021Auxiliary} use the composed query image retrieval as an auxiliary task to enhance the training process, which can be seen as a self-supervised method. \cite{yan2021l2c} employ extra general image captioning data to improve semantic understanding.
However, none of these works attend to fully explore the interaction across modalities and take most advantage of the given data. Instead, we propose a new pre-training and finetuning schema to align visual and textual representations at a fine-grained level.

\noindent \textbf{Vision-Language Pre-training }
Recently, Vision and Language Pre-training (VLP) methods~\cite{lu2019vilbert, chen2020uniter, li2020unicoder, li2020oscar,zhou2020unified, Kim2021ViLTVT, huang2021seeing, hu2021vivo} have shown their success on multi-modal tasks by learning cross-modal representations. 
However, these pre-trained models are not applicable to the IDC task as they lack the ability of comparing, and the learned representations are too coarse. We follow these works to tailor self-supervised tasks for IDC to enhance the cross-modal alignment and make best use of the given data.
Moreover, several VLP related works~\cite{radford2021learning,li2020unimo,huo2021wenlan,lee2021umic} introduce contrastive learning  to unify different modal representations. 
The major idea is to pull closer the distances between positive image-text pairs and push away negative pairs.
Motivated by these works, we introduce Contrastive Learning to our pre-training tasks and construct negative samples in different granularity, which enhances more fine-grained cross-modal alignment.

\section{Conclusions}

In this paper, we propose a new pretraining-finetuning paradigm for image difference captioning (IDC), which can align visual differences and textual semantics at a fine-grained level. We specifically design three self-supervised tasks with contrastive learning strategies to learn stronger association across modalities. Experiments on CLEVR-Change and Birds-to-Words demonstrate the effectiveness of our proposed pre-training tasks. To address the limitation of supervised IDC data, we explore to utilize two cross-task datasets on Birds-to-Words and prove that they can provide more background knowledge. Our proposed model architecture is flexible to accommodate more data of different forms. In the future work, we will further explore automatic construction of large-scale feasible data to enhance the pre-training stage and improve model generalization ability.

\section*{Acknowledgment}
This work was partially supported by the National Natural Science Foundation of China (No. 61772535 and No. 62072462) and Beijing Natural Science Foundation (No. 4192028).

\bibliography{reference}

\appendix
\clearpage
\section{Appendix}


\subsection{More Implementation Details}
\label{details}
Table~\ref{tab:paras} presents the detailed hyper-parameter setups during pre-training and fine-tuning. 
We train the model on GPU 1080ti. We fix all the random seed in experiments to 1234.

\begin{table}[htb]
\begin{tabular}{l|l|l}
\toprule
Hyper-parameters & CLEVR-Change & Birds-to-Words \\
\midrule
Max sent len     & 40           & 80             \\
Layer num        & 3            & 2              \\
SE block         & 1            & 1              \\
PE block         & 1            & 1              \\
Hidden size      & 512          & 512            \\
FFN hidden size  & 2048         & 2048           \\
Attention heads  & 8            & 8              \\
Warmup Steps     & 8K           & 4K             \\
Training steps   & 250K         & 50K            \\
Learning rate    & 1e-4         & 1e-4           \\
Weight decay     & 0.0          & 0.0            \\
Dropout          & 0.2          & 0.2            \\
Batch size       & 200          & 200            \\
Training time    & 66h          & 20h           \\
\midrule
\midrule
Learning rate    & 3e-5         & 3e-5           \\
Dropout          & 0.2          & 0.2            \\
Batch size       & 200          & 32             \\
Weight decay     & 0.0          & 1e-4           \\
Main metric      & CIDEr        & ROUGE-L        \\
\bottomrule
\end{tabular}
\caption{Hyper-parameters used in pre-training and finetuning stages. The parameters for pre-training and for finetuning are shown in the first and second blocks respectively. SE block denotes the number of transformer layers in the single-image encoder and PE block refers to the number of transformer layers in the pair-image encoder.}
\label{tab:paras}
\end{table}

\begin{figure}[htb]
    \centering
    \includegraphics[width=\linewidth]{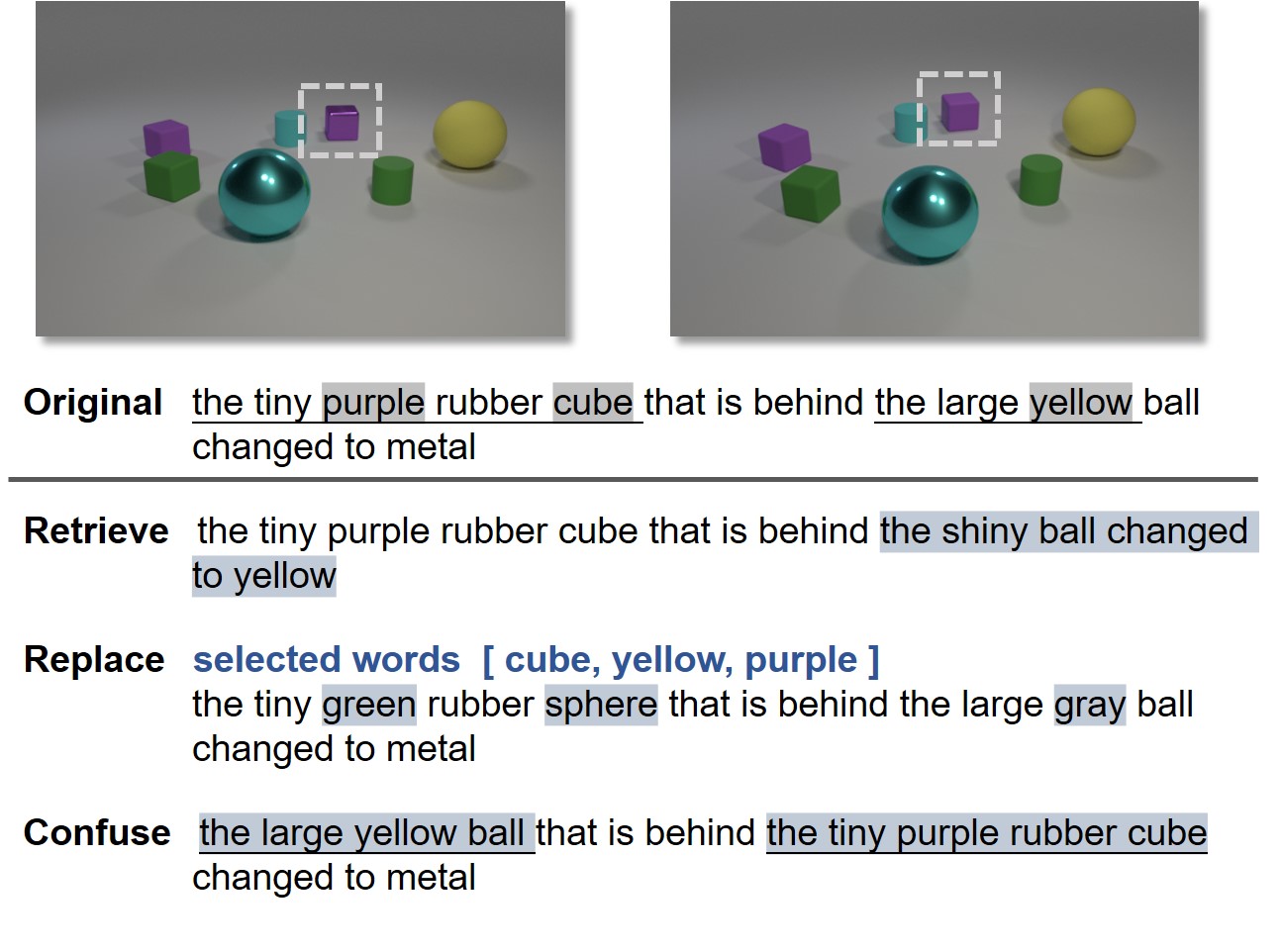}
    \caption{An example of constructed negative sentences for the FDA task on CLEVR-Change Dataset.}
    \label{fig:clevr_negs}
\end{figure}


\subsection{Negative Samples Constructed in FDA}
We construct three types of negative samples in FDA task including \textit{Retrieve}, \textit{Replace}, and \textit{Confuse}.  The automatically constructed negative samples on CLEVR-Change is shown in Figure~\ref{fig:clevr_negs}. 
In the \textit{Replace} strategy, we select top 50\% most important nouns and adjectives in the sentence and replace them with other words
having the same POS-Tags from
pre-defined vocabularies. The statistics of average replaced words and the average sentence length are represented in Table~\ref{tab:replace}. We pre-define  POS-Tags vocabularies following the steps: 1) annotate all nouns and objectives in the sentences of training set; 2) select the top 50\% most important annotated words in each sentence to obtain vocabularies of different POS-Tags; 3) filter the words whose document frequency is less than 10. The pre-define  vocabularies of Birds-to-Words is shown in Table~\ref{tab:vocabs}. We construct 6 negative sentences for each image pair in the triplet, that is positive:negative=1:6.

\begin{table}[htb]
\begin{small}
\begin{center}
\begin{tabular}{l|l|l|l}
\toprule
Dataset        & \#avg words & \#avg len & \%   \\
\midrule
CLEVR-Change   & 2.5              & 10.8          & 23.4 \\
Birds-to-Words & 5.8              & 32.4          & 18.0 \\
\bottomrule
\end{tabular}
\end{center}
\caption{Statistics of \textit{Replace} strategy in FDA on two benchmark datasets. \textcolor{black}{\textbf{\#avg words} refers to the average number of replaced words in a sentence, \textbf{\#avg len} refers to the average sentence length, \textbf{\%}  refers to the average percentage of words replaced in a sentence.}}
\label{tab:replace}
\end{small}
\end{table}

\begin{table}[htb]
\begin{small}
\begin{center}
\begin{tabular}{l|c|l}
\toprule
POS-Tags & Size & Words       \\                                \midrule      
JJ       & 187  & ``white'',``long'',``light'' \dots     \\
JJR      & 49   & ``bigger'',``thinner'',``darker'' \dots \\
NN       & 158  & ``beak'',``breast'',``stripe'' \dots    \\
NNS      & 73   & ``legs'',``wings'',``feathers'' \dots  \\
\bottomrule
\end{tabular}
\end{center}
\caption{Pre-defined vocabularies of Birds-to-Words dataset.}
\label{tab:vocabs}
\end{small}
\end{table}

\begin{figure*}[htb]
    \centering
    \includegraphics[width=0.65\linewidth]{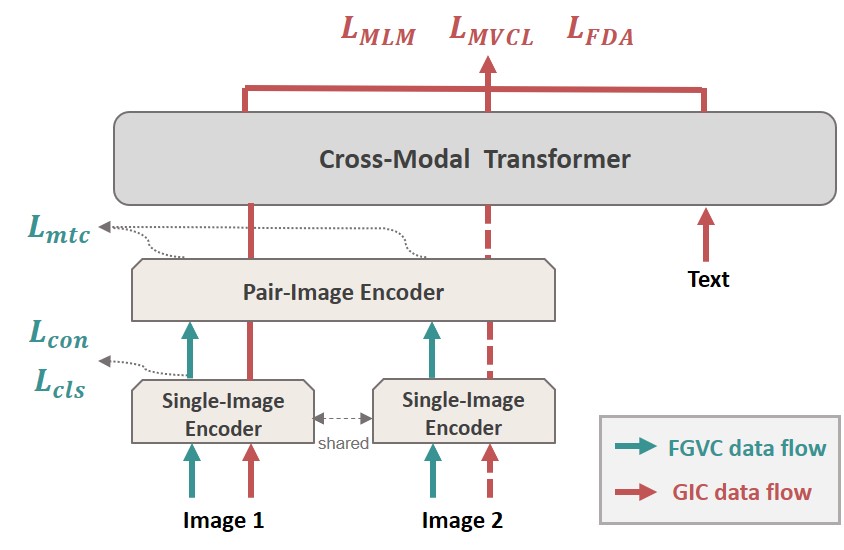}
    \caption{Visualize of cross-task data expansion strategy. We utilize extra data from general image captioning (GIC), that is the CUB dataset, and Fine-grained visual classification (FGVC), that is the NABirds dataset. For the data flow of FGVC, $\mathcal{L}_{\text{cls}}$, $\mathcal{L}_{\text{con}}$, $\mathcal{L}_{\text{mtc}}$ denotes classification loss, contrastive learning loss and matching loss respectively. For the data flow of GIC, $\mathcal{L}_{\text{MLM}}$, $\mathcal{L}_{\text{MVCL}}$ and  $\mathcal{L}_{\text{FDA}}$ are the objective of the three pre-training tasks.}
    \label{fig:in_domain}
\end{figure*}

\subsection{Details of Cross-task Data Usage}
Considering that the number of training samples in the Birds-to-Words dataset is limited but the semantics are quite diverse, we apply additional cross-task datasets to alleviate the problem of limited data. To be specific, we utilize a general image captioning dataset CUB and a fine-grained visual classification dataset NABirds. How they are used or the data flow of these two datasets in our architecture is illustrated in Figure~\ref{fig:in_domain}. The FGVC data (NABirds) is used to enhance the image difference encoder. Specifically, we construct  48,562 image pairs. The single-image encoder is refined by a fine-grained classification loss $\mathcal{L}_{\text{cls}}$  and a  contrastive loss $\mathcal{L}_{\text{con}}$, which pulls the image representations of the same label in a batch closer and pushes image representations of different labels farther apart. The pair-image encoder is refined by a matching loss $\mathcal{L}_{\text{mtc}}$. Meanwhile, the GIC data (CUB) can provide image-text pairs and promote alignment between image and text representations. We draw dashed lines to represent the second pseudo image (the empty image) of the image-pair in GIC. The pseudo triplets from GIC can be applied to the three pre-training tasks. In the MVCL task, we only mask the real image. In the FDA task, since there is no difference description in the general image caption dataset, we do not apply \textit{Confuse} in negative samples construction. We construct negative samples  via \textit{Retrieve} and \textit{Replace} only based on the GIC data. 


We use the two datasets in two stages of pre-training. The CUB dataset is used in the first  stage to initialize the model parameters and learn primary alignment between image and text. The NABirds dataset is used together with Birds-to-Words in the second stage. The batch ratio of  NABirds and Birds-to-Words is 1:4.

\begin{figure*}[htb]
    \centering
    \includegraphics[width=0.8\linewidth]{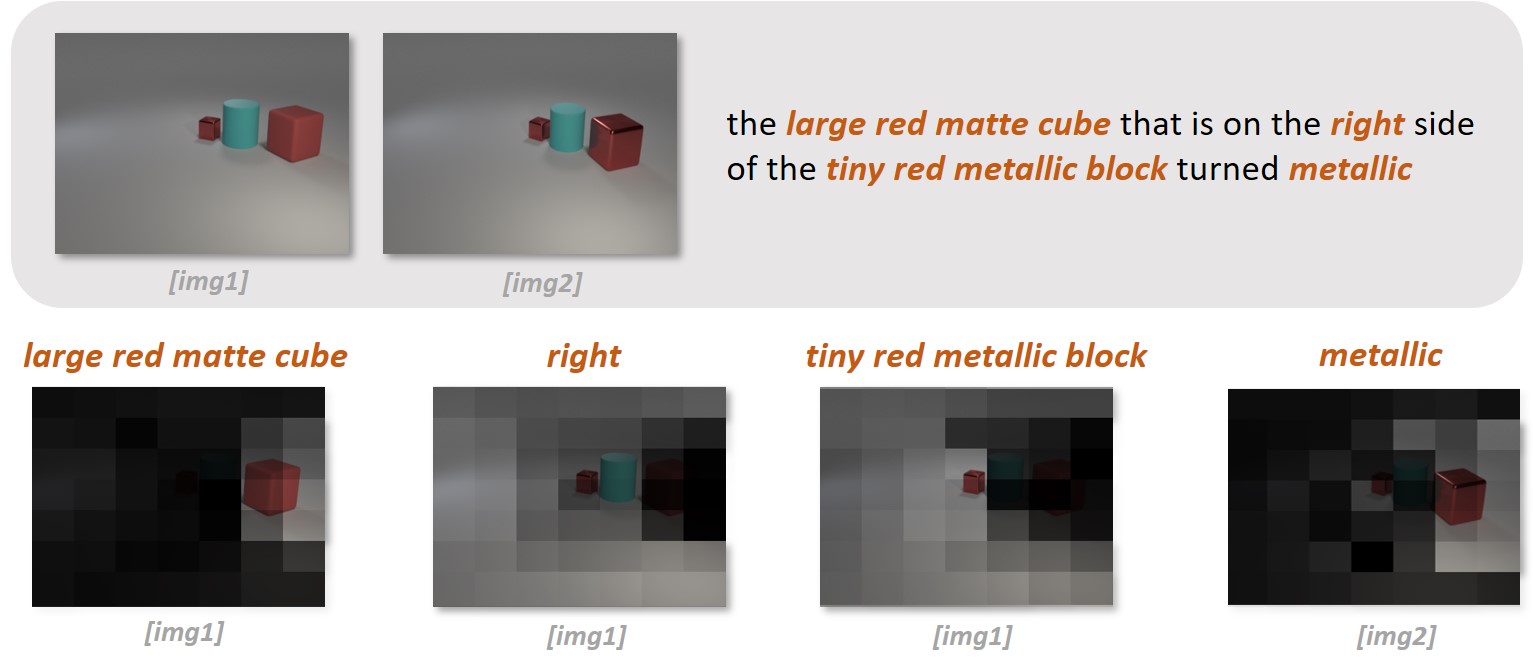}
    \caption{Visualizations of cross-modal alignment  on CLEVR-Change dataset.}
    \label{fig:align}
\end{figure*}

\begin{figure*}[htb]
    \centering
    \includegraphics[width=\linewidth]{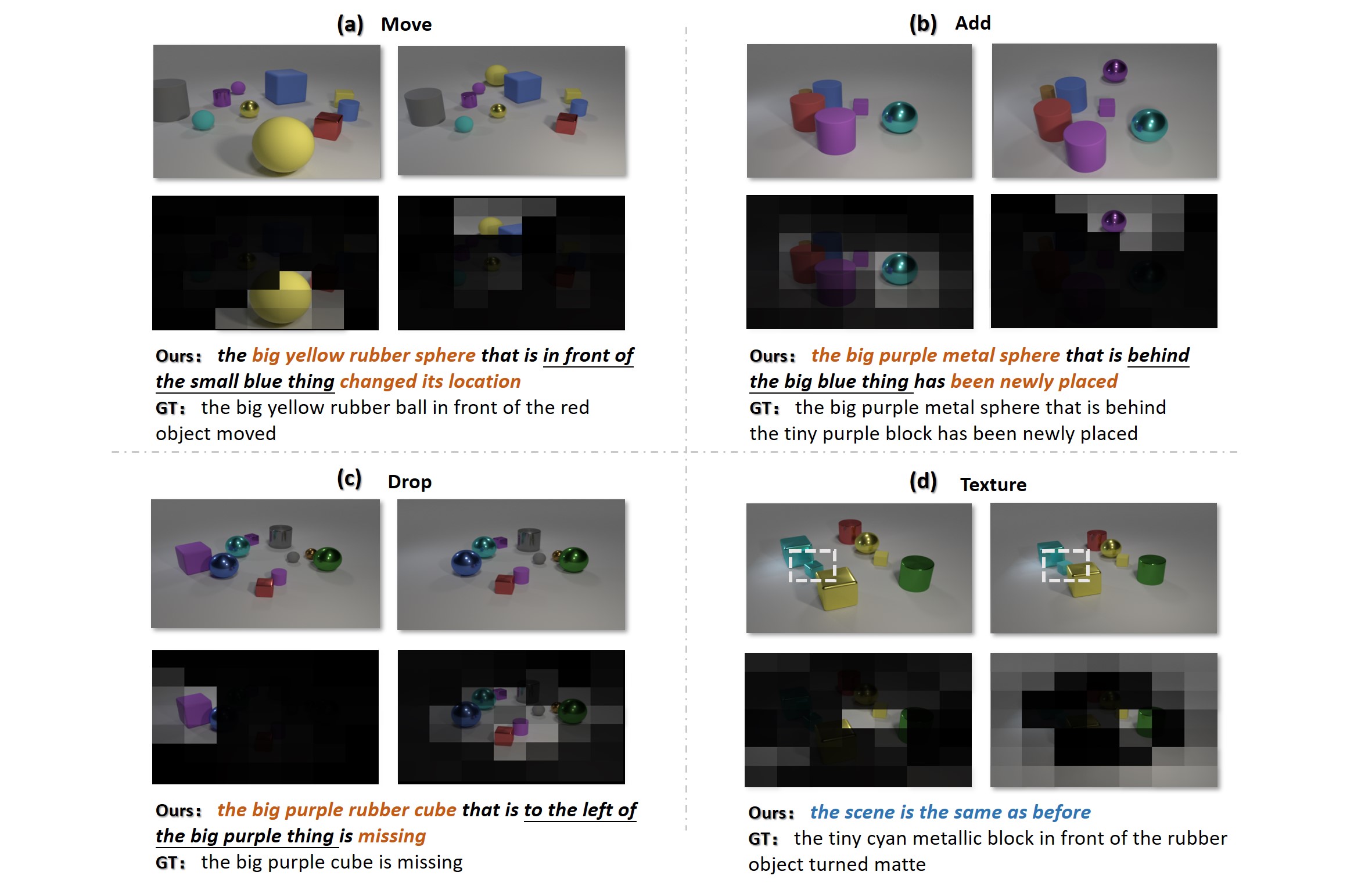}
    \caption{Visualizations of four generated cases on CLEVR-Change dataset of different change types, including: (a) Move an object, (b) Add an object, (c) Drop an object, (d) Change an object's texture. Our model can accurately capture different types of fine-grained changes and locate the change regions as shown in cases (a)-(c). However, as shown in case (d), it fails to locate the fine-grained texture change of a tiny object as the changes between the two images are somewhat complicated, including view shift and object  occlusion in addition to the texture change.}
    \label{fig:clevr_case}
\end{figure*}

\begin{figure*}[htb]
    \centering
    \includegraphics[width=\linewidth]{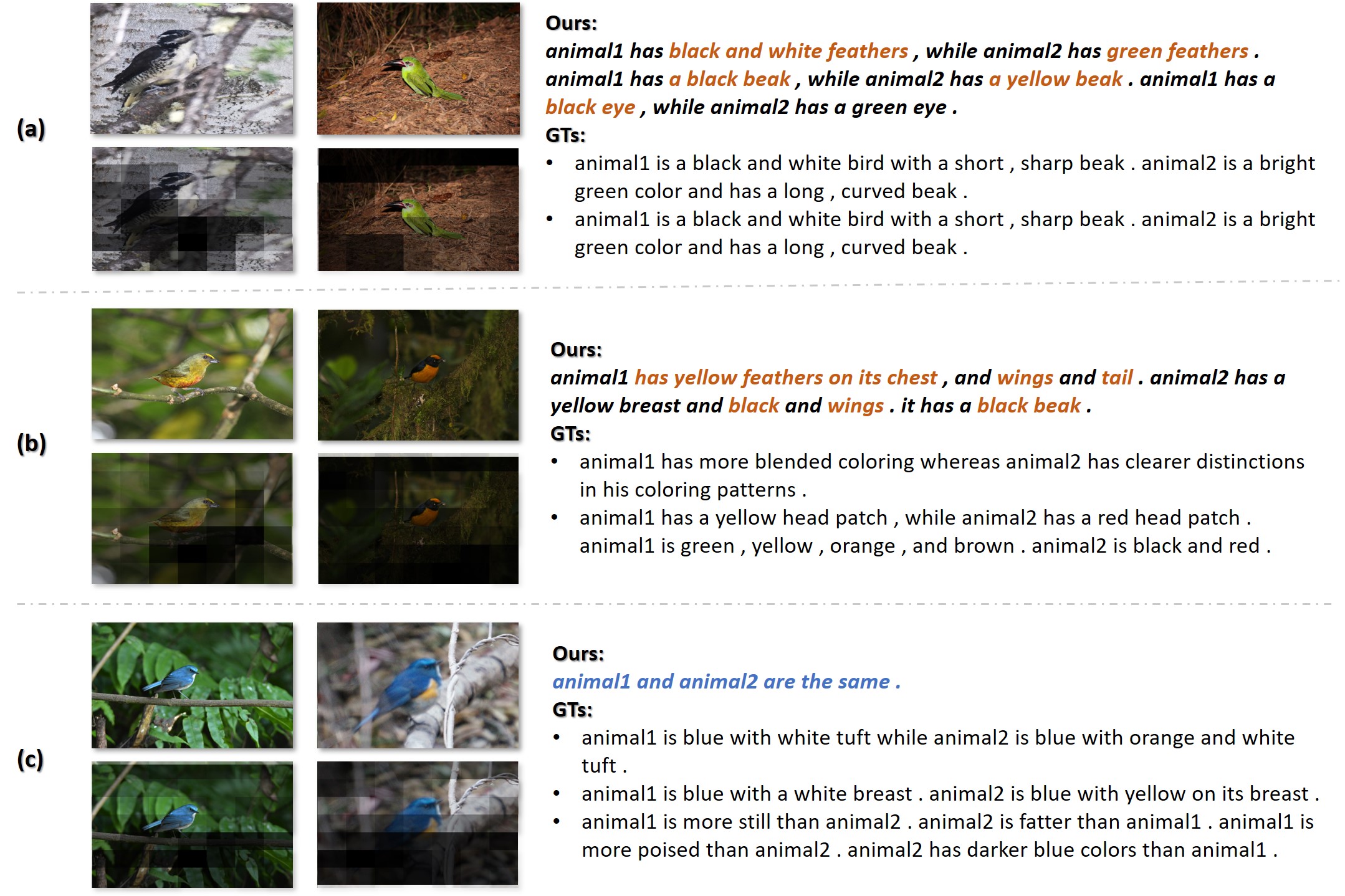}
    \caption{Visualizations of several generated cases on Birds-to-Words dataset. The orange bold words are correct difference-related words.}
    \label{fig:bird_case}
\end{figure*}


\subsection{Visualization and Analysis}
\vspace{10pt}
\noindent \textbf{Cross-Modal Alignment}
To demonstrate that our proposed model can align the visual difference with the textual descriptions at a fine-grained level, we visualize the text-to-image attention maps of the cross-modal transformer in Figure~\ref{fig:align}. An unseen triplet sample from the test set is input to the pre-trained model to show the cross-modal alignment.

\vspace{10pt}
\noindent \textbf{Case Study}
We present more generated difference descriptions in Figure~\ref{fig:clevr_case} and  Figure~\ref{fig:bird_case}. To be specific, Figure~\ref{fig:clevr_case} illustrates four different types of changes in CLEVR-Change: (a) Move an object, (b) Add an object, (c) Drop an object, (d) Change an object's texture. The first three cases show that our model can handle the  different type of fine-grained changes and locate the change region accurately. We particularly present a case (d) with wrong difference description.\textcolor{black}{ The model fails to locate the fine-grained texture change of a tiny object, because the changes are somewhat complicated, including view shift and object occlusion in addition to the texture change.} 

Figure~\ref{fig:bird_case} shows three difference description cases on the Birds-to-Words dataset. The first two cases indicate that our model can handle various fine-grained bird appearance in the complex environment. Compared with human annotations, the proposed model may miss some details of bird species, as shown in case (c). It is more challenging to spot and describe the visual differences in this dataset, since the bird species have diverse attributes and the language of the annotated descriptions is richer.

\end{document}